# Bose-Einstein Condensation of Photons in a Microscopic Optical Resonator: Towards Photonic Lattices and Coupled Cavities

Jan Klaers, Julian Schmitt, Tobias Damm, David Dung, Frank Vewinger, and Martin Weitz
Institute for Applied Physics, University of Bonn, Wegelerstr. 8,
53115 Bonn, Germany

Bose-Einstein condensation has in the last two decades been observed in cold atomic gases and in solid-state physics quasiparticles, exciton-polaritons and magnons, respectively. The perhaps most widely known example of a bosonic gas, photons in blackbody radiation, however exhibits no Bose-Einstein condensation, because the particle number is not conserved and at low temperatures the photons disappear in the system's walls instead of massively occupying the cavity ground mode. This is not the case in a small optical cavity, with a low-frequency cutoff imprinting a spectrum of photon energies restricted to values well above the thermal energy. The here reported experiments are based on a microscopic optical cavity filled with dye solution at room temperature. Recent experiments of our group observing Bose-Einstein condensation of photons in such a setup are described. Moreover, we discuss some possible applications of photon condensates to realize quantum manybody states in periodic photonic lattices and photonic Josephson devices.

## 1. INTRODUCTION

When a gas of particles is cooled, or its density is increased up to the point that the associated de Broglie wavepackets spatially overlap, quantum statistical effects come into play. Specifically, for material particles with integer spin (bosons), the phenomenon of Bose-Einstein condensation into a single quantum state then minimizes the free energy. Bose-Einstein condensation was first experimentally achieved in 1995 by laser and subsequent evaporative cooling of a dilute cloud of alkali atoms [1,2]. The condensate atoms can be described by a macroscopic single-particle wavefunction, similarly as known for liquid helium [3]. Other than material particles, photons usually show no Bose-Einstein condensation [4]. Thermal photons commonly have no chemical potential, corresponding to a non-conserved particle number upon temperature variations, and in blackbody radiation photons at low temperature vanish in the cavity walls, instead of exhibiting a phase transition to a macroscopically occupied ground state mode. Early theoretical work has proposed the Bose-Einstein condensation of photons in Compton scattering of X-rays in plasmas [5]. After the realization of atomic Bose-Einstein condensates [1,2], we have witnessed an increased interest in light sources, where a macroscopically populated photon mode is not the result of laser-like gain, but rather stems from a thermal equilibrium phase transition. In a laser, both the state of light field and that of the amplifying, usually inverted, medium is very far removed from thermal equilibrium [6]. Lasing is even a prominent example for a non-equilibrium process: only the absence of thermal equilibrium allows for inversion and optical gain. In interesting work, Chiao et al. proposed a two-dimensional quantum fluid in a nonlinear Fabry-Perot resonator [7]. Thermal equilibrium was sought from photon-photon scattering, similarly to atom-atom scattering in atomic physics BEC experiments, but the limited photon-photon interaction in available nonlinear materials has so far prevented a sufficient thermalization [8]. In other work, the demonstration of (quasi-) equilibrium Bose-Einstein condensates of exciton-polaritons [9,10], mixed states of matter and light, and magnons [11], has been achieved. In the former experiments, interparticle interactions of the excitons; that is the material part of the polaritons, drive the system into or near thermal equilibrium. More recently, evidences for the superfluidity of polaritons have been reported [12,13].

The experiments described here build upon dye molecules in liquid solution. Thermalization of a photon gas inside an optical resonator with the dye solution is achieved by repeated absorption and reemission processes in the dye. The excited dye molecules here can, in the sense of a grand canonical ensemble, be understood as a reservoir, which exchanges both energy and particles with the photon gas. For such systems it is known that frequent collisions (~$10^{-14}$ s timescale) of solvent molecules prevent coherent energy exchange between photons and dye molecules, so that the strong coupling regime is not fulfilled and the experiment is well described by the uncoupled degrees of freedom, photons and molecules respectively [14,15].

We here describe recent experiments of our group demonstrating Bose-Einstein condensation of photons in a dye-solution filled optical microresonator [16]. The photons here much act like a gas of material particles, with an observed phase transition temperature, which is at room temperature, being many orders of magnitude higher than in an atomic physics Bose-Einstein condensate. A further notable system property is that loading and the thermalization of photons, by radiative coupling to the dye molecules, proceeds throughout the experiment time, i.e. the system can constitute a continuous output condensate. In this article we moreover discuss some possible future studies of the novel two-dimensional photonic condensate in detail, in the presence of interactions and external potentials. In a periodic potential, the Mott-transition for the photon gas can be investigated, and more complex entangled manybody states. Interestingly, quantum manybody states, when constituting the system ground state, are expected to be populated directly by cooling into the multiparticle state in our system, an issue that has not been achieved in ultracold atomic gas systems.

## 2. PRESENT EXPERIMENTS

In our work, a dye-filled optical microresonator is used, in which photons are frequently absorbed and reemitted by the dye molecules (Fig. 1a) [16,17]. The small distance between the two spherically curved mirrors of 3.5 optical wavelengths causes a large frequency spacing between adjacent longitudinal cavity modes. The latter is comparable with the emission width of the dye molecules, see Fig. 1b. In combination with an intracavity modification of the spontaneous emission, preferring the emission to small volume modes (low transverse excitation), a regime is reached, where to good approximation the resonator is populated only with photons of a fixed longitudinal mode, q=7 in our case. With the longitudinal mode number frozen the system is effectively two-dimensional. Moreover, the photon dispersion becomes modified with respect to free space and becomes quadratic, that is particle-like. The frequency of the transverse $TEM_{00}$-mode here is the lowest allowed mode, and acts as a low-frequency cutoff. One further finds that the mirror curvature leads to harmonic confinement of the photon gas (diagram on the left of Fig. 1a). In general, we expect significant population for the $TEM_{nm}$ modes with high transversal quantum numbers n and m (of high eigenfrequency) at high temperature, while the population concentrates to the low transverse modes when the system is cold.

Equilibrium is reached as photons are absorbed and emitted by dye molecules many times. This leads to a thermal population of cavity modes with a temperature determined by the (rovibrational) temperature of the dye molecules, which is at room temperature. During thermalization (in an idealized experimental situation) only transversal modal quantum numbers are varied, and, different than in the blackbody radiator, no photons are destroyed or created in average. This is because thermal emission of photons is negligible in the limit of a cutoff frequency in energy units ($\hbar\omega_{cutoff} \cong 2.1$ eV for light in the yellow spectral regime) being far above the thermal energy (1/40 eV at room temperature). The dye molecules have quantized electronic excitation levels with transition energy near or above the cutoff, whose thermal excitation is suppressed by a factor of order $\exp(-\hbar\omega_{cutoff}/k_BT) \sim 10^{-36}$. One can show that the description of the photon gas in the resonator in the paraxial limit is formally equivalent to that of a two-dimensional harmonically trapped gas of massive particles with effective mass $m_{eff}= \hbar\omega_{cutoff}/c^2$. For a harmonically trapped two-dimensional ideal gas it is known that a Bose-Einstein condensate exists at finite temperature, in contrast to the two-dimensional homogeneous case [18]. Thus, a BEC is expected at sufficiently low temperature and high density.

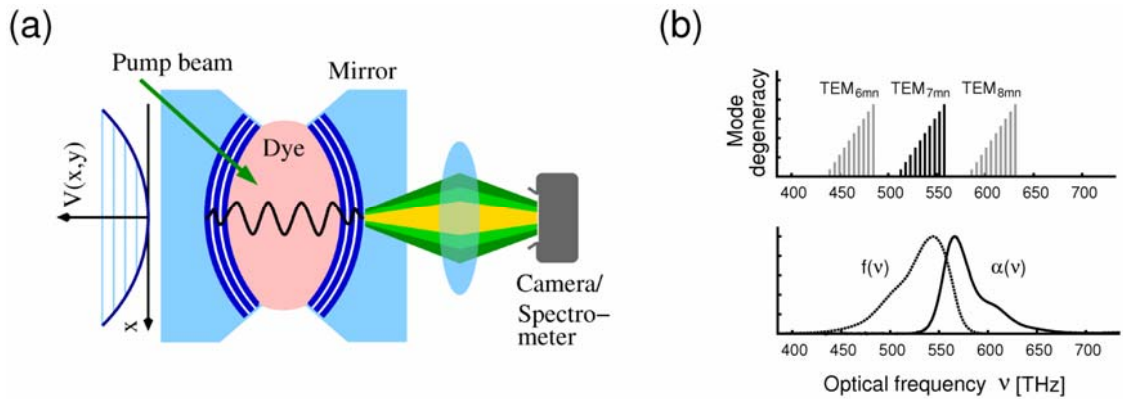

Fig. 1: (a) Scheme of the experimental setup. The trapping potential for photons imposed by the curved mirrors is indicated on the left hand side. (b) Schematic spectrum of cavity modes. Transverse modes belonging to the manifold of longitudinal mode number q=7 are shown by black lines, those of other longitudinal mode numbers in grey. The bottom graph indicates the relative absorption coefficient and fluorescence strength of rhodamine 6G dye versus frequency.

Thermalization of photons trapped in the resonator proceeds by multiple absorption and emission processes, resulting in thermal contact with the dye solution. This stems from a thermalization of the rovibronic dye state [19]. Consider an idealized dye molecule with an electronic ground state $S_0$ and electronically excited state $S_1$, each subject to additional rovibronic splitting, as shown in Fig. 2a. Frequent collisions of solvent molecules with the dye molecules, on the timescale of a few femtoseconds at room temperature, rapidly alter the rovibrotional state of the dye. These collisions are many orders of magnitude faster than the electronic processes (the dye upper state lifetime is a few nanoseconds), so that both absorption and emission processes can well be assumed to take place from a state where the rovibrotional manifold is in thermal equilibrium with the dye solvent. One can show that the spectral distributions of absorption $\alpha(\omega)$ and emission $f(\omega)$ respectively will be linked by a Boltzmann factor $f(\omega)/\alpha(\omega) \sim \omega^3 \exp(-\hbar\omega/k_BT)$, a relation long known as the Kennard-Stephanov relation [19] (which is valid both for spontaneous and stimulated emission respectively). By multiple absorption and emission processes, as indicated in Fig. 2b, the thermalized distribution is transferred to the spectral distribution of the photon gas. In the experiment, the thermalization occurs as the photon cycles many times back and forth between the resonator mirrors. This drives the system towards a thermal distribution of modes above the cavity low-frequency cutoff [17].

Other than in a perfect system, in a real experiment photon loss by the finite dye quantum efficiency, mirror losses and coupling modes unconfined in the optical resonator occur. In our setup, these are compensated for by pumping the dye medium with a laser beam of wavelength near 532 nm. Thus, by external pumping we generate and uphold a reservoir of electronic excitations in the dye that can effectively exchange particles (and energy) with the photon gas. In this model, the photon gas is described by a grandcanonical ensemble, with the dye acting as a reservoir. The particle exchange with the reservoir is of special relevance for the expected photon statistics of the condensate [20,21]. Experimentally, we observe, despite losses and pumping, a spectral photon distribution that in steady state is very close to an equilibrium distribution [17]. We attribute this to the thermalization process proceeding sufficiently fast, i.e. the photons scattering several times off a molecule before being lost.

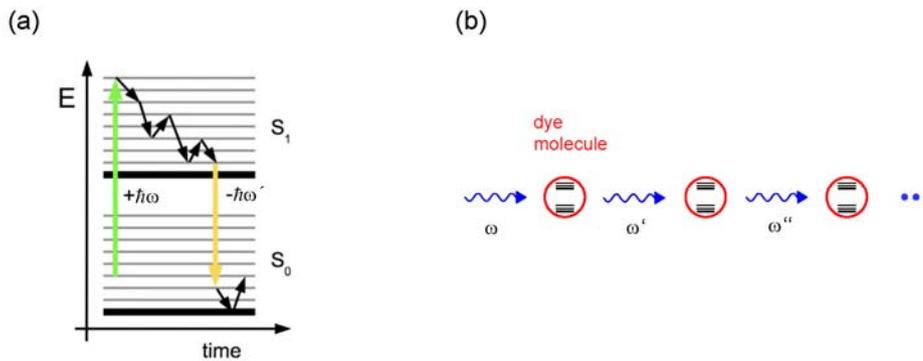

Fig. 2: (a) The rovibrational structure both in the electronic ground ($S_0$) and excited ($S_1$) state of dye molecules thermalizes by frequent scattering of solvent molecules. (b) Scheme of the thermalization of photons by repeated absorption and emission processes. The photon gas in this way thermally couples to the dye solution reservoir, resulting in a thermal distribution of modes above the cavity low-frequency cutoff.

In initial experiments, we have carefully tested for thermalization of the two-dimensional photon gas in the dye-filled microresonator [17]. In this low density regime, well below the threshold to a BEC, the chemical potential is strongly negative. Moreover, a spatial concentration was observed, which is a consequence of the thermalization within the effective trapping potential, formed by the curved mirrors. This behaviour is evident from the analogy with a trapped gas of massive particles, e.g. atoms, at finite temperature, and we expect that this effect offers prospects for the spatial concentration of diffuse sunlight to a central spot.

In subsequent experimental work, we have succeeded in the observation of a Bose-Einstein condensate of photons, for which an increased optical pumping rate was required. Fig. 3a shows typical spectra of the photon gas at different optical pump rates [16]. While the observed spectrum at small intracavity optical powers resembles a Boltzmann-distribution above the low-frequency cutoff, near the expected phase transition a small shift towards the cutoff frequency is seen, and the spectrum more resembles a Bose-Einstein distribution. At optical intracavity powers above the critical value, in addition a spectrally sharp peak at the position of the cutoff is visible, the Bose-Einstein condensate. The observed spectral width of the condensate peak is limited by the resolution of the used spectrometer. The experimental results are in good agreement with theoretical expectations (see the inset of the figure). At the phase transition, the optical intracavity power is $P_{c,exp}=(1.55 \pm 0.6)$ W, which corresponds to a photon number of $(6.3 \pm 2.4) \cdot 10^4$.

Fig. 3b shows spatial images of the radiation transmitted for one of the cavity mirrors (real image onto a colour CCD camera) both below (top) and above (bottom) the condensate threshold. Both images show a shift from the yellow spectral regime for the transversally low excited cavity modes located near the trap center to the green for transversally higher excited modes appearing at the outer trap regions. In the lower image, a bright spot is visible in the center with a measured FWHM diameter of $(14 \pm 2)$ μm. This corresponds well to the expected diameter of the $TEM_{00}$ transverse ground state mode of 12.2 μm, yielding clear evidence for a macroscopic population of the ground state mode. Fig. 3c gives normalized intensity profiles (cuts along one axis through the trap center) for different powers. We observe that not only the height of the condensate peak increases for larger condensate fractions, but also its width, see also Fig. 3d. This effect is not expected for an ideal photon gas, and suggests a weak repulsive self-interaction mediated by the dye solution. The origin of this is most likely thermal lensing, but in principle it could also be due to a Kerr-nonlinearity in the dye medium. On a mean-field level, both effects can be modelled by an optical self-interaction of the light field (see [16]), and by comparing the observed increase of the mode diameter with numerical solutions of the two-dimensional Gross-Pitaevskii equation, we estimate a dimensionless interaction parameter of $\tilde{g} \approx (7 \pm 3) \times 10^{-4}$. This is significantly smaller than the values $\widetilde{g} = 10^{-2} \ldots 10^{-1}$ reported for two-dimensional atomic physics quantum gas experiments [22,23] and also below the values at which Kosterlitz-Thouless physics can be expected to become important in the harmonically trapped case [24]. An indication for this would be the loss of long-range order. Experimentally, when directing the observed peak through a Michelson-type sheering interferometry, we have not seen signatures for the phase blurring that was observed in two-dimensional atomic gas experiments [22]. This supports the conclusion that indeed a BEC is observed.

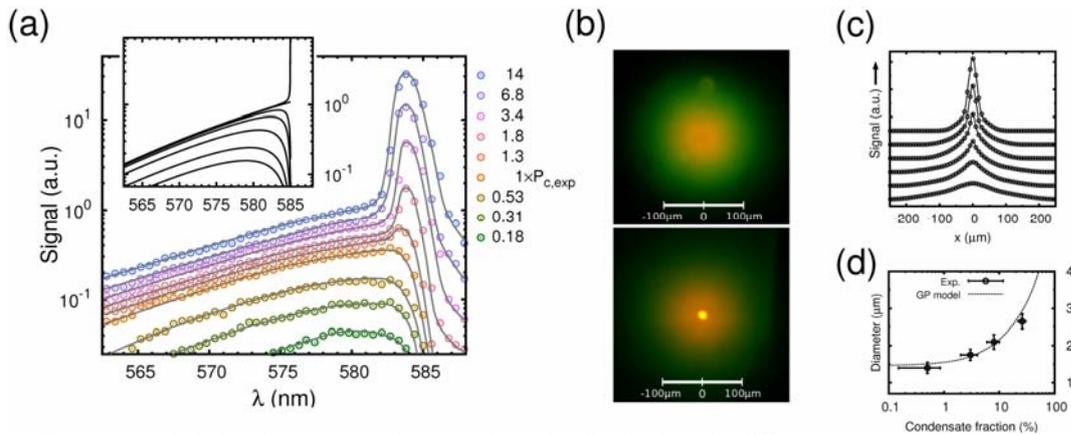

Fig. 3: (a) The connected circles show measured spectral intensity distributions for different pump powers. The legend gives the optical intracavity power, determining the photon number. On top of a broad thermal wing, a spectrally sharp condensate peak at the position of the cavity cutoff is visible above a critical power. The observed peak width is limited by the spectrometer resolution. The inset gives theoretical spectra based on Bose-Einstein distributed transversal excitations. (b) Images of the radiation emitted along the cavity axis, below (top) and above (bottom) the critical power. In the latter case, a condensate peak is visible in the center. (c) Cuts through the center of the observed intensity distribution for increasing optical pump powers. (d) The data points give the measured width of the condensate peak versus condensate fraction and the dotted line is the result of a theoretical model based on the Gross-Pitaevskii equation.

Further evidences for the observation of a photon Bose-Einstein condensate include the phase transition setting in both at the expected critical photon number and showing the expected scaling with resonator geometry, and the ground state mode emerging even for a spatially displaced pump spot, as can be understood from the thermalization [17]. In more recent work, we have also investigated thermalization of the photon gas by replacing the dye solution with semiconductor quantum dot nanocrystals, with which also signatures for a thermalisation could be observed [25].

Recently, we have theoretically investigated the expected condensate fluctuations of the photon gas system [21]. Due to the possible effective particle exchange with the excited state dye molecules, the light condensate represents a grandcanonical Bose-Einstein condensate, for which enhanced particle fluctuations, as compared to an atomic Bose-Einstein condensate or also laser, are expected. Measurements on the photon statistics of the dye-filled microcavity system are currently performed, and will be reported elsewhere.

## 3. PERIODIC POTENTIALS FOR THE PHOTON GAS

### 3.1 Optical lattices

We now discuss the possibility to create periodic potentials for the photon gas in the dye-filled microcavity system. In the field of ultracold atomic gases, the use of periodic potentials from the ac Stark shift induced by off-resonant laser fields is a well established technique [26]. In the strongly correlated regime, a phase transition between the superfluid and a Mott-insulating phase, with unity occupation of atoms per site, has been observed [27]. The strong confinement reachable in optical lattices has also allowed for the achievement of low-dimensional systems, where new phases can be present. For two-dimensional atomic gases, a crossover to a Kosterlitz-Thouless phase with quasi-long-range order was observed [22]. Two-dimensional quantum gases are also of large interest for fractional quantum Hall type physics, and this is one goal of current research with rapidly rotating atomic Bose-Einstein condensates [26]. For early theoretical work considering phase transitions in lattices with photonic systems see [28,29].

Lattice-type structures have been investigated experimentally in exciton-polariton systems. An interesting work reported the realization of a one-dimensional condensate array formed by depositing metal strips onto a semiconductor microcavity structure [30]. For this system, the spontaneous built-up of states with a relative phase relationship between neighbouring lattice sites of 0° in phase and 180° out of phase, respectively, has been demonstrated. In other experiments with a double-well polariton system, Josephson oscillations have been observed [31]. We are also aware of long standing work on Josephson physics in superconducting systems, and work in ultracold atomic gases [3,32,33].

To achieve a spatially periodic trapping potential for the two-dimensional photon gas, a corresponding periodic refractive index variation of the dye solution in a direction transverse to the resonator axis has to be introduced. The modulation of the refractive index can be achieved by applying standing wave patterns with additional optical lattice beams which locally heat the dye or induce a microscopic Kerr nonlinearity. In the former case, local heating leads to thermo-optic nonlinearities. For example, a second dye species with low quantum efficiency and an absorption resonance spectrally shifted from the wavelength of the photons confined in the resonator could be added to the dye solution. Irradiation with an optical field tuned to the absorption resonance of this "heating" dye will now locally, via the resulting temperature increase, lower the refractive index, and thus at the corresponding transversal position lessen the optical length between mirror surfaces. One can readily show that this is equivalent to the application of a local potential increase for the photon gas in the paraxial limit, as can be understood from the smaller optical wavelength (corresponding to a higher photon energy) required to locally match the mirrors' boundary conditions.

In this fashion, a transverse standing wave pattern allows for the thermo-optic imprinting of a spatially periodic potential for the photon gas, see Fig. 4. To create sufficiently high potential wells at small lattice spacing, we estimate that a local temperature change of the dye of a few Kelvin is required. An alternative solution for the creation of optical lattices would be to replace one of the cavity mirrors by a mechanical grating-type structure. Along similar lines, photonic crystal manufacturing technologies could be used to create periodic optical length modulations within the cavity, which for the photon gas translates to an effective, spatially periodic potential. The challenge in the two described alternative approaches is to maintain a sufficiently high cavity finesse.

For a shorter spacing between sites, as achieved with a spacing between sites of slightly over a micron for typical potential well depths in our configuration, tunneling of the photon gas between sites becomes important. In principle, the interplay between tunneling and mean-field interactions of the photon gas from thermo-optic self-interactions allows for the transition to a state in which the number fluctuations per site are reduced, as compared to a Bose-Einstein condensed state. Probably the most interesting experiment that can be carried out in this regime of mean-field interactions is with a double-well potential and a large number of photons per well, which then constitutes a Josephson system for the photon gas. Josephson physics has previously been investigated the context of both ultracold atomic gases and exciton polariton condensates [31-33]. In the dye microcavity system, a double-well system could be implemented by imprinting a thin potential well within the microcavity, cutting the condensate region in half, and we expect that both Josephson oscillations and the self-trapping regime can be observed.

Further, by admixing solvents with large Kerr effect, as nitrobenzene or diacetylene-polymer materials, rapid variations of the refractive index could be obtained to tailor the photon interactions on a microscopic, nondissipative level, where the interactions are equivalent to effective photon-photon scattering [8]. The interplay of tunnelling and interactions in this microscopic regime is well described by a Bose-Hubbard model for photonic gas, which could allow for an observation of the Mott-insulator transition for the photon gas in the lattice.

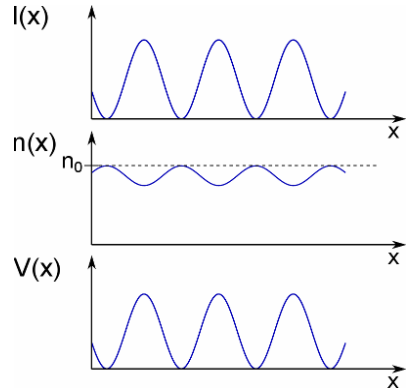

Fig. 4: Schematic illustration of proposed generation of periodic lattice potential for the photon gas. The irratiation of the optical lattice beams near the standing wave maxima reduces the index of refraction of the dye solution (top and middle), and shortens the optical distance between the mirrors for the trap wavelength. This induces an effective repulsive potential for the photon gas, whose spatial variation (bottom) follows the lattice beams standing wave pattern.

Other than in a cold atomic gas system, the cooling (to room temperature) by the fluorescence induced thermalization process proceeds throughout the experimental manipulation cycle, and in the presence of loading (i.e. with irradiation by the pump laser beam).

**3.2 Photon gas thermalization in coupled cavity systems**

We next discuss the possible preparation of a macroscopic occupation of entangled photon states through condensation using coupled microcavities. We are aware of ongoing work studying coupled optical microcavity systems in the context of quantum information, where the configurations are tailored to minimize the coupling to the environment [34-36]. This is highly complementary to the here described approach, where cavities filled with dense media for thermal coupling of the photonic state to the room temperature environment will be used. Fig. 5 gives a schematic of a system with two dye-filled microcavities, in each of which a two-dimensional periodic potential is created for the investigated photon gas by standing wave patterns in the dye medium through interfering laser beams. The cavity number that a photon populates acts, besides photon polarisation and lattice site number, as an additional degree of freedom for the photon. Coupling of photons is again achieved by effective photon-photon interactions in the dye solution. In the coupled system, entangled multiparticle states can become the relevant energy eigenstates of the system. The thermalization now proceeds towards a population of the lowest eigenstate. We expect that a macroscopic population of such entangled states by condensation can be achieved in such a coupled cavity system. While a usual BEC consists of a macroscopic accumulation of single-particle states, in the coupled cavity system we expect to be able to create a product of entangled multi-particle states. The corresponding scheme represents a novel approach for quantum state preparation, with an entangled target state being populated in a thermal equilibrium process, when it constitutes the system ground state.

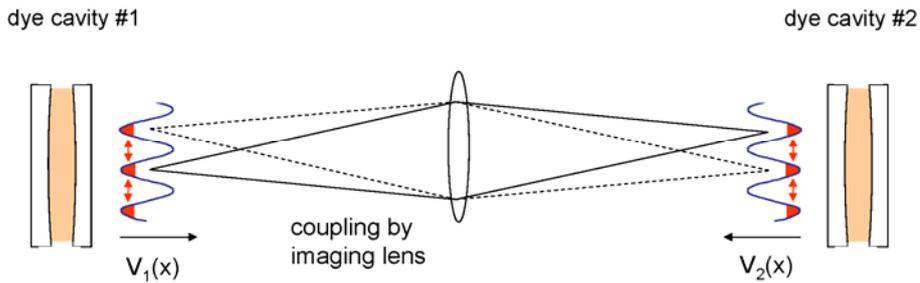

Fig. 5: Two coupled dye-filled microcavities for condensation into a macroscopic occupation of entangled multiparticle states. In each of the cavities, external laser beams (not shown) imprint a periodic trapping potential for photons. The red arrows indicate a tunnel coupling between neigbouring trapping sites, and photons within a single site interact through e.g. the Kerr nonlinearity. Coupling between sites in different cavities is achieved with a high numerical aperture optical imaging lens. For a coupled system of two Josephson junctions, one would replace the lattice potentials by double well potentials respectively.

An alternative approach for the generation of entangled states by thermalization is the use of a coupled system of Josephson junctions for the photon BEC operating with mean-field interactions alone, an approach that is motivated by work in superconducting systems [3]. Similar as in the above described approach, this again requires the nonclassical target state to be the system ground energy eigenstate in a coupled cavity system. Further perspectives include the realization of more complex many-body states in the two-dimensional quantum gas. Examples are topological states, as can be realized in a system with three coupled cavities and using a Berry phase. This can allow for the study of correlated quantum Hall type states in the photon BEC system by thermalization alone, when this highly nonclassical state constitutes the system ground state.

## 4. CONCLUSIONS

To conclude, we have described current work and future perspectives of photonic quantum gases in optical microcavities under thermal equilibrium conditions. To date, Bose-Einstein condensation of atomic gases achieves a macroscopic population of single particle states, while the generation of manybody states requires subsequent, in most experiments coherent, manipulation steps. In contrast, photonic quantum gases, with effective interactions in a medium and under suitable confinement, can provide a route towards the realisation of a macroscopic accumulation of entangled states directly through condensation. It is argued that this allows for an efficient and straightforward preparation of entangled states via a thermal equilibrium process.